\documentclass[twocolumn,showpacs,preprintnumbers,amsmath,amssymb,prl]{revtex4}

\usepackage{graphicx}
\usepackage{dcolumn}
\usepackage{bm}

\begin{document}


\title{The Innsbruck EPR Experiment:\\ %
          A Time-Retarded Local Description of Space-like Separated Correlations}

\author{Michael Clover}
\email{michael.r.clover@saic.com}
\affiliation{%
Science Applications International Corporation\\
San Diego, CA  }%

\date{\today}

\begin{abstract}
A local, time-retarded hidden variable model is described that fits the recently measured EPR data from the Innsbruck collaboration. The model is based on the idea that entangled waves in the zero-point field convey information from the detectors to the source, stimulating the spontaneous emission of unentangled photons with definite polarizations.  In order to match experimental data, the model is augmented with a further local assumption that the ``master'' photon (going back along the direction of the zero-point wave that triggered emission) will not be detected if the polarizer is not in the same orientation as the stimulated wave.  This model predicts a ratio of coincidences to singles of 1/3 compared to standard quantum mechanics' 2/3 for the 2-fold choice of modulator settings in this experiment, and predicts that a 20-fold choice of  settings will yield a coincidence ratio of 1/40 with the SQM ratio unchanged.  Such an outcome should be easily distinguishable given the  Innsbruck group's measured efficiency of 1/20 in their experiment.  This model also predicts that a coincidence at polarizer settings $(\vec{\alpha},\vec{\beta},t)$ will never be correlated with its time-retarded complement setting, $(\vec{\alpha}',\vec{\beta}',t-2L/c)$, a prediction  for whose test  the limited data publicly available at this time is inadequate. 
\end{abstract}

\maketitle

\section{\label{sec:background}Background}

If one takes  locality seriously,  recognizing that reality is best described by the equations of QED and QFT rather than non-relativistic Schr\"odinger equations, then one would expect any explantion of EPR experiments to involve time-retardation effects mediated by  zero-point fields.

It does not seem to be widely known or appreciated that one can describe the process of spontaneous emission within QED as the stimulated emission by the zero-point ($\frac{1}{2} \hbar \omega$) waves, although to be correct, the zero-point waves account for  only half the spontaneous emission and the radiation reaction the other half, when time-symmetric ordered operators are used~\cite{milonni}.  

Given this justification, and assuming that if standard quantum mechanical photons can maintain their entanglement through kilometers of optical fiber then the zero-point field can also support entanglement over those lengths,  we assume the stimulation of ``spontaneous'' emission is related to time-retarded polarizer positions.  We further assume that the polarization of emitted photons is not isotropic but only in the directions of the polarizers (and $90\deg$ offset).  That is to say, we use the zero-point field to collapse the wave-function at the instant of emission instead of at the instant of detection.

\subsection{\label{sec:proof_loopholes} Loopholes in the Proofs}

Most derivations of the Bell/CHSH inequalities~\cite{B64,CHSH69,CH74}, in making the assumption of locality, assume that the hidden variable ($\lambda$) is not an explicit function of the detector settings (at the time of detection).  In  footnote 13(b) of reference~\cite{CH74}, Clauser and Horne explicitly reject the possibility that ``systems originate at the analyzers and impinge upon the source, thus effecting the ensemble in a manner dependent upon $a$ and $b$.''  This dependence of $\lambda$ on the detector settings at the retarded time, of course, is exactly what obtains if the time-retarded zero-point field induces emission, vitiating their derivation.

Aspect~\cite{A76} calculated a  correlation coefficient that was dependent upon the time-retarded detector settings, but by assuming that each possible retarded state was an independent random variable, his constraint doesn't apply to the current model of dependent correlated variables.

The proofs of non-locality are not that strong.

\subsection{\label{sec:experimental_loopholes}Experimental Loopholes}

If the various photon detectors are not perfectly efficient (5-10 percent are recent numbers~\cite{WZ98, KZ95}), then standard quantum mechanical   (SQM) predictions may {\textit{not}} violate the ``strong'' Bell inequalities.  Only by making assumptions of  ``joint detection independence''~\cite{CHSH69}, ``no enhancement''~\cite{CH74} or ``faithful sampling''~\cite{AGR82, WZ98} can one derive weaker Bell inequalities that SQM does violate. 

In the case that photon detectors don't detect every photon, various authors~\cite{ GG02, CG94, SZ00} have noted that local models can be constructed that reproduce the quantum mechanical violations of the weaker Bell inequalities.  These constructs are {\textit{ad hoc}}, and no attempt is made to provide a physical justification for the features of the model(s).  Interestingly, the mere possibility that both photons from a parametric down-conversion event might end up in a single detector lowers the quantum mechanical predicted values to within the Bell inequality bounds~\cite{ CG94}.

\section{The Time-Retarded Model}

The   algorithm is quite simple: given a coincidence at time $t$ in  detectors at distance $L$ from a source, it posits that the source emitted particles  at  time $t- L/c$, based on stimuli from the detector/polarizers at time $t-2L/c$ (stimuli that may have traveled down long fiber optic cables).  In a static EPR setup, this zero-point-stimulated ``spontaneous'' emission comprises  4 polarities of equal weight (intensity) in any unit of time: aligned or not with Alice's polarizer at $\alpha$ or $\alpha^{\perp}=\alpha + \frac{\pi}{2} $, or aligned or not with Bob's polarizer at $\beta$ or $\beta^{\perp}$.    In essence, this picture implies that when a polarizer is introduced to an experimental setup, the character of spontaneously emitted photons changes from uniformly distributed polarization vectors to  nonuniform fixed directions, keeping the overall intensity constant.

\subsection{A Static Experimental Interlude}

There are two general types of static EPR experiments:  those using a polarizer followed by a single detector, and those using polarizing beam splitters (PBS's) followed by two detectors.  

In the former case, this model has 50 percent efficiency:  the quarter of  photons emitted at angle $\alpha$ are detected byAlice's detector with perfect efficiency and its mates in Bob's  detector with probability $\sim \sin^2(\alpha - \beta)$; similarly for the quarter emitted at $\beta$.  The other quarters, at $\alpha^{\perp}$ and $\beta^{\perp}$, are absorbed by the polarizing media.

In the latter (PBS) case, $\alpha$'s are detected perfectly  (in the ($+$) channel) and all the mates are detected, split between Bob's two channels with probabilities $\cos^2(\alpha - \beta)$ and $\sin^2(\alpha - \beta)$; ditto for the $\beta$'s.  The $\alpha^{\perp}$'s and  $\beta^{\perp}$'s are similarly detected  in the respective ($-$) channels,  and their mates split similarly. 

In either such  static experimental setup, this algorithm clearly reproduces the standard quantum mechanical predictions,  the only (conceptual) difference being that the wavefunction has ``collapsed'' at the moment of emission instead of at the moment of detection.

\subsection{The Dynamical Experiment of  Weihs, {\textit {et al.}}~\cite{WZ98}}

In  the Innsbruck experiment~\cite{WZ98}, the experimental setup used  dynamic polarization modulators in front of static polarizing beam-splitter detectors.   Alice's  two-channel polarizer was aligned at the angles  $\alpha=\frac{\pi}{8}$ and  $\alpha^{\perp}= \frac{5\pi}{8}$, while Bob's was aligned  at $\beta= 0$ and $\beta^{\perp} =\frac{\pi}{2}$.  The modulator in front of each detector was adjusted to either perform no modulation, or to rotate the polarization by $\frac{\pi}{4}$  from $\alpha$  to  $\alpha'$ or $\beta$ to  $\beta'$.  The choices to modulate by $0$ or  $\frac{\pi}{4}$ were made in a  random manner on a time-scale of order 100 ns, significantly less than the time of flight of the photons.

 The experiment measured coincidence rates between the various combinations of transmitted($+$) and reflected($-$) beams in the two detectors.  Standard quantum mechanics predicts that 
 $C^{qm}_{++}(\alpha, \beta) \propto \sin^2(\beta - \alpha)$,
 since the parametric down-conversion process results in the two photons' polarizations at right angles to each other.

    The coincidence rates are combined into an expectation value for that setting:
 \begin{eqnarray}
 E(\alpha,\beta) = \frac{C_{++} + C_{--}- C_{+-} - C_{-+}}{C_{++} + C_{--}+ C_{+-} + C_{-+}} \ ,\nonumber
 \end{eqnarray}
($E^{qm}(\alpha,\beta) = -\cos2(\beta-\alpha)$) and these are combined into the Bell/CHSH parameter:
 \begin{eqnarray}
S(\alpha,\alpha',\beta,\beta')& =  & |E(\alpha,\beta)  - E(\alpha',\beta) | \nonumber \\
                            &&+ |E(\alpha,\beta')  + E(\alpha',\beta')  | \ . \nonumber
 \end{eqnarray}
 For local realistic models that conform to the derivations of the inequality, $S \le 2$, while  $S^{qm} \le 2\sqrt{2}$.
 
 The rapid polarization modulation does have the effect of putting Alice's choice of modulator setting outside the light cone of Bob's detector for any given measurement. However, these detector settings see a common source and that source in its turn will only have seen a finite number of different detector settings in its past light cone:  at its moment of emission, it will have been stimulated  by  $(\alpha,\beta)$, $(\alpha,\beta')$, $(\alpha',\beta)$, or $(\alpha',\beta')$.  By virtue of the complete randomness of the choices, each of these possibilities should occur one quarter of the time. 

\subsection{The Original Algorithm}

Our first model makes no further assumptions about the behavior of polarized photons:  if  an $\alpha$-polarized photon hits a polarizer at angle $\alpha$, it is  transmited with unit probability; if  it encounters a polarizer at angle $\alpha'$, it is transmited with probability $\cos^2(\alpha-\alpha')$.  The associated photon in the other direction has similar transmission probabilities (e.g. $\cos^2(\alpha^{\perp} -\beta)$). Coincidence probabilities multiply, and we average the four possible emission patterns onto the current setting of the detector to cumulate a correlation function.

\begin{figure}
   \begin{minipage}[t]{1.0\linewidth}  
    \includegraphics[width=\textwidth]{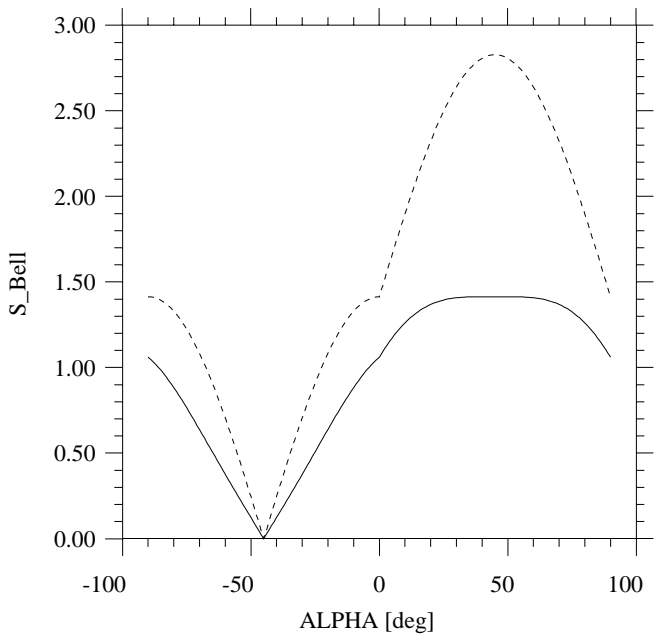}
    \caption{\label{fig:sbell1} The Bell/CHSH parameter for various amplitudes of Alice's polarization modulator, using the original local model. The local  model (solid) obeys  Bell's inequalities ($S\le  \sqrt{2}\  ({\textit{i.e.}}\ \le 2)$) while the SQM result (dashed) satisfies $S\le 2\sqrt{2}$.  The local and SQM coincidence/singles ratios are equal.}
  \end{minipage}
\end{figure}

The Innsbruck experimenters also did an experiment where they varied the voltage on Alice's modulator, in order to measure other angles than the four ``canonical'' ones.  Figure~\ref{fig:sbell1} shows the results that this local model predicts for the Bell parameter for different modulator settings.  It is clear that this model never violates the Bell inequality, while the SQM results  do.  We remark that the total coincidence rates of the various detectors ($N={C_{++} + C_{--}+ C_{+-} + C_{-+}}$) were identical for the  local and SQM models -- both had perfect efficiency -- and that both were independent of the modulator amplitude (half the individual rates showed sinusoidal behavior, as expected).

\subsection{The Second Local Model}

Our second model introduces the notion of a ``master'' photon, in order to recover the SQM coincidence pattern.  If the source emits a ``master'' photon with polarization $\alpha$ which encounters a polarizer in configuration $\alpha$, then the photon will always be detected and its slave  projected as usual onto $\beta$.  On the other hand, if the ``master'' (at $\alpha$)  encounters the polarizer in configuration $\alpha'$,  that photon will not be detected at all, but its slave will contribute to the singles' count rates in the usual manner. In this  aspect, this model resembles the prism model of Fine~\cite{SZ00}; it differs in that for a static experiment, this model has perfect efficiency. 

\begin{figure}
   \begin{minipage}[b]{1.0\linewidth}  
      \includegraphics[width=\textwidth]{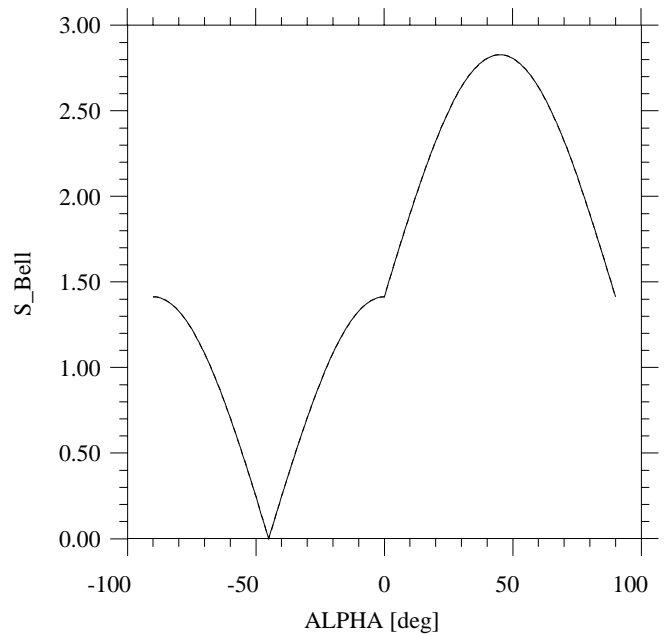}
      \caption{\label{fig:sbell2} The Bell/CHSH parameter for various amplitudes of Alice's polarization modulator. The local and SQM values overlie each other.  The local coincidence/single ratio is half the SQM ratio.}
  \end{minipage}
 \end{figure}

This model's predictions for the Bell parameter are shown in figure~\ref{fig:sbell2}, where they overlie the SQM prediction for all values of the polarization modulator amplitude. 

 The total coincidence rates for this experiment with this model, as one might have expected,  are always half of those of the SQM prediction.  Given that half of all the PDC events double up in one fiber or the other (for this experiment's fiber coupling), the ratio of coincidences to singles for this model is 1/3, while SQM would expect to see 2/3.  Weihs {\textit{et al.}}~\cite{WZ98} note that their experiment  only observed a ratio of about $0.05$ given the  efficiencies of their detectors. 
 
  If each modulator had had three settings, this local model would have required the non-detection of $2/3$ of the ``master'' photons, and in general, $(n-1)/n$ for $n$ settings.  An experiment whose modulators could randomly choose between 20 different amplitude settings  could differentiate between this model (1/40) and SQM (2/3) at the current  detector efficiencies (1/20).  
  
 This model  makes a further prediction that can also be tested with less than ideal detectors:  if Alice and Bob  examine the polarizer patterns that occur $2L/c$ earlier than each occurrence of, {\textit{e.g.}} an $(\alpha,\beta)$ coincidence, they should never find an $(\alpha', \beta')$ occurrence; all other autocorrelation times should see that pattern in one quarter of the histories, assuming the times are binned in units of the polarizer modulator period (100 ns in this experiment), although in principle one would prefer to see the retarded settings' time  ($-2L/c$)  measured with the same temporal acceptance window used for  coincidences between Alice and Bob.

 \subsection{Mining the Innsbruck Data}
 
 After downloading the sample data provided on the web by the Innsbruck group~\cite{WZweb}, a program was written to examine the coincidence data and look at temporal correlations.  Alice's 388K and Bob's 302K singles events  yielded 15134 coincidences in a $6$ ns window.  These events generated a Bell parameter of  $2.71 \pm 0.50$ \footnote{The error of $\pm 0.50$ results from assuming statistical independence of the  16 terms contributing  to $S_{Bell}$,  adding the relative errors in quadrature, given that  the individual $C_{ij}$'s were between $186$ and $1992$ for this data set. }, consistent with the published result of $2.73 \pm 0.02$.

   Of these, there were  only 19 coincidences separated by a time interval between 4.3 and 5.2 $\mu$s.  Table~\ref{tab:coin_pat} indicates how these 19 were distributed in 3 coarse bins centered around the round-trip travel time, $ 2L/c =  4.8 \mu$s
 \footnote{Given that the fibers were 500m long, and that at 700nm the index of refraction in fiber cable is about 1.45  then $c = 20.7 $ cm/ns, and  $2L/c \sim 4.8 \mu$s.}.  Unfortunately, the statistical uncertainty in these numbers is so large that no definitive conclusion can be drawn -- one would like to see about 100 events in  6 ns windows about $2L/c$ for this experiment.  

 \begin{table}
\caption{\label{tab:coin_pat}Distribution of Alice/Bob settings at time intervals near $2L/c$. ``Both'' means that both Alice and Bob's polarizer settings were the same for the given interval time, ``one'' means that one of Alice or Bob's settings was changed, and ``neither'' means that both settings were different.  This last is ruled out by the time-retarded local model. The lower rows show how these detector patterns are distributed over larger time interval bins.}
\begin{ruledtabular}
\begin{tabular}{cccccc}
$\Delta t$ [$\mu$s] &Total & both  & one  & neither & ratio (n/T)\\
\hline
4.3 - 4.6 & 2 & 1 & 0 & 1 & 0.50\\
4.6 - 4.9 & 6 & 0 & 4 & 2 & 0.33\\
4.9 - 5.2 & 11 & 0 & 6 & 5 & 0.45\\
\hline
$0$        - $10^1$  & 171     & 40    & 65       & 66       & 0.39    \\
$10^3$ - $10^4$ &  206k & 53k & 103k     & 51k    & 0.25   \\
$10^6$ - $10^7$ & 9.3M & 2.4M & 4.6M & 2.3M  & 0.29  \\
\end{tabular}
\end{ruledtabular}
\end{table}

\section{Conclusion}

We have presented a  time-retarded local  hidden variable theory that ``collapses the wave-function'' at the source rather than at the detector.  This model  can reproduce  all  statically measured correlation experiments,  with perfect efficiency.  It can also reproduce the  dynamically space-like separated correlation data of Weihs {\textit{et al.}}~\cite{WZ98}, albeit with 50 percent efficiency. 

 We have found that there is insufficient data available to determine whether  there is  an absence of ``complement'' patterns of Alice and Bob's polarizer settings at a retarded time of $2L/c$  in accordance with this time-retarded local model and propose further data-mining and experiments.  We also propose what should be a practical experiment involving multi-choice ($\sim 20-$way) polarizers that should be able to discriminate between this local model and SQM with current detector efficiencies.

Most importantly,  this counter example shows that the various derivations of Bell's inequalities, all of which assume a time-independent ``locality'',  do not apply to any of the experiments conducted to date with static (or stochastically periodic) settings.  It is not enough that Alice and Bob make decisions independent of each other, those decisions cannot be the same as those made $2L/c$ earlier, if experiments are to preclude local processes mediated by zero-point waves.

The ``non-localities'' of 3- and 4-particle GHZ experiments also  appear to be explainable with time-retarded zero-point waves that  stimulate the sources to emit radiation polarized or phased in exactly the correct manner to be detected in the various setups, {\textit{e.g.}}~\cite{RT90}.

\begin{acknowledgments}
We wish to acknowledge Lewis Little~\cite{LL94}, who started us thinking about how information could get from detectors back to sources.
\end{acknowledgments}

\bibliography{epr}

\end{document}